# Pipeline Leak Detection Techniques

**Assist.Prof. Timur Chis, Ph.D., Dipl.Eng.**
**"Andrei Saguna" University, Constanta, Romania**

ABSTRACT. Leak detection systems range from simple, visual line walking and checking to complex arrangements of hardware and software. No one method is universally applicable and operating requirements dictate which method is the most cost effective. The aim of the paper is to review the basic techniques of leak detection that are currently in use. The advantages and disadvantages of each method are discussed and some indications of applicability are outlined.

## 1. Introduction

Our need to transport fluids from the point of production to the area of end use has led to a rapid increase in the number of pipe lines being designed and constructed. Many of these carry toxic and hazardous products, often close to centers of high population or through areas of high environmental sensitive. With the need to safeguard these lines, on-line monitoring is becoming routine and in some case 24 hours surveillance is mandatory. With the increase in Rumanian pipeline deterioration by thief and terrorism, the need for rapid and reliable pipe line measurement and control systems will increase.

The review begins with a summary of causes of leaks and the implications of the failure to detect them. Basic techniques are covered and the features of each are briefly discussed. The bulk of the paper deals with modern computer based techniques. Basic flow equations are covered and the on-line dynamic calculation required are listed together with the impute data required to enable the monitoring to be accomplished with the minimum of downtime. The latest systems are capable of resolving down to 1% of maximum rated flow with a response time of a few minutes. Practical experience verifies this figure but the total cost of such a system could be high. The system therefore becomes a compromise, between response, performance, and alarm availability, and choice of instrumentation is crucial. The integration of good quality instruments with advanced real time models seems to be the current trend and the paper closes with some personal thoughts on future trends.

There are four main categories of pipe line failures. These are:
-pipeline corrosion and wear;
-operation outside design limits;
-unintentional third party damage;
-intentional damage.

Many pipelines are operated for a number of years with no regard to any possible mechanical changes occurring in the line. Some of the products may be corrosive, the pipe line may be left partially full for periods of time, or atmospheric effects may cause external damage. These three reasons are responsible for pipe line corrosion and this may give rise to corrosion "pits" developing along the line. These are small in nature and could be responsible for material imbalances over a period of time. Very accurate flow metering can be used to detect this as discussed in the next section. Abrasive fluids or dust-laden gas streams can give rise to pipe line weir. Again, this is a slow process, but should a weak spot develop (more often than not close to a change in direction or section) then a pipe break may occur very rapidly and totally unexpectedly.

Operation outside design guidelines is more common than is realized, as operators seek to use the line for as many fluids as possible. If the line is deigned for a certain maximum temperature and pressure, then operation at higher pressure and/or high temperature could lead to spontaneous failure. The problem could be compounded if the line has a large but unknown amount of corrosion. Unintentional third party damage may occur if excavation or building occurs near buried lines. More often than not the right-of-ways are not clearly marked and lines are sometimes broken by bulldozers or similar plant machinery possibly with fatal results.

Intentional damage unfortunately is on the increase and pipe line carrying flammable or high value products make ideal targets. Alarm systems linked to block valves can help to minimize the amount of products reels as a result of sabotage, so again certain lines are instrumented with the intention of reducing the effects of planned terrorism.

The cost of failure to detect leaks also falls into four main areas:
-loss of life and property;
-direct cost of loss products and lie downtime;
-environmental cleanup cost;
-possible fines and legal suits.

These are all self explanatory, with the most costly of these being the last, although any of four areas could be very expensive. The size of claims can run into many millions of dollars, so the cost of fitting and operating leak detection systems is often insignificant compared with the cost of failure of the line.

It is this background that is causing operators and designers to turn to on-line program integrity monitoring systems.

## 2  SIMPLE LEAK DETECTION SYSTEMS

The most basic method of leak involves walking, driving, or flying the pipe line right-of-way to look for evidence of discoloration of vegetation near the line or actually hear or see the leak. Often unofficial pipe line monitoring is performed by people living nearby who can inform the operator of a problem with the line.

The most cost effective way to detect leaks in non-flammable products is to simply add an odorant to the fluid. This requires some care in selection, as frequently the odorant has to remove before the transport fluid can be used. Organic compounds make the most useful odorizes, especially when the fluid being carried has no natural smell of its own. The disadvantage of such method is that if the leak occurs in an area of no population the leak will go undetected unless the line is walked regularly by pipe line surveillance crews carrying suitable "stiffer" detector. Thus to the apparent low cost of this method have to be added the cost of removing the odorant and maintaining staff to check the line at frequent intervals. The location of a leak is also dependent on prevailing weather conditions. Strong winds may disperse the smell and atmospheric inversion may give an incorrect location of leak and the uncertainty of relying on this method alone is high. Nevertheless it is useful method if used in conjunction with other techniques.

Simple line flow balances are frequently used to check from gross imbalances over hourly or daily based. This method may identify that a leak is present but flow meters at each end of the line will not identify the leak location. A line pressure measurement system will be required in conjunction with the flow meters to establish that the pressure gradient has changed from the on-leek situation. The method is useful, however in identifying the existence of corrosion pits as the outputs of flow meters at each end of his line will consistently diverge if flow in the line is maintained constant. If line flow rate varies with time, that imbalances are more difficult to detect, since the flow meter outputs ay vary nonlinearly with flow rate or may have different flow characteristics from each other.

A loss of product will be identified simply as the difference between the steady state inventory of the system and the instantaneous inlet and outlet flows. Mathematically this is:

$$\Delta V = V_{in} - V_{out} - V_l \qquad (1)$$

where:

$\Delta V$ =leakage volume;
$V_{in}$=meter inlet flow;
$V_{out}$=meter outlet flow;
$V_l$=pipe line fluid inventory.

This last term can be calculates As the average of the integrate inlet and outlet flows in simple systems, but as will be seen later, the value of this term can be calculated more accurately and easily in real time as a function of several variables.

Another method is based on detecting the noise associated with or generated by a leak. There are many instances where fluid flow can generate vibration at frequencies in excess of 20 Hz. These frequencies are in the ultrasonic range but can be detected with suitable transducers. The device can be made portable so that pipe line crews can clamp a transducer at any point along the line to check for noise by noting the signal strength, the source of the leak can be pinpointed.

A similar technique, though base on a different principle, is the acoustic „wave alert" monitor, more correctly called a negative pressure wave detector. This a piezoelectric sensor that give an output went dynamically stressed. When a leak followed by rapid line repressurization a few milliseconds later. The low pressure wave moves away from the leak in both directions at the speed of sound. The pipe walls act as a waveguide so that this rarefaction wave can travel for great distances attenuating in amplitude as if does so.

Sensors placed at distances along the line can be triggered as the wave passes and location of the leak can be calculated from the line conditions and the internal timing devices in the instrument.

Such devices are particularly useful in identifying large breaks in lines very rapidly since the transient wave typically moves 1 mile in 5 seconds in gases and almost 1 mile per second in liquids. Response is therefore on the order of a few seconds depending on the positioning of the transducers.

## 2. PIG BASED MONITORING SYSTEMS

Pipe lines are frequently used for pipe line commissioning, cleaning, filling, dewaxing, batching, and more recently pipe line monitoring. This last type of pig can be designed to carry a wide range of surveillance and monitoring equipment and can be used at regular intervals to check internal conditions rather than continuously monitoring the line. Data, however, can be built up over a period of time to provide a history at the line. This information can be used to predict or estimate when maintenance, line cleaning, or repairs are required. If a leak is detected, for example, by flow meter imbalance, the location can be found by using a pig with acoustic equipment on board. This will alarm when the detection equipment output reaches a maximum and the precise location of pig can be confirmed by radio transmitters also mounted on board. Pigs require tracking because they may become stuck, at a point of debris build-up. Pigging should be carried out at a steady speed, but occasionally the pig may stop and start, particularly in smaller lines. Information on when and where the pig stop is therefore important in interpreting the inspection records. Pig tracking is not new and many such proprietary systems exist. In the best systems, however, a picture of the line is often programmed in so that outputs from junctions, valves, crossovers ad other geometries act as an aid to location. Pig tracking can make use of the acoustic methods discussed earlier. When the sealing cups at the front of the pig encounter a weld, vibrational or acoustic signal are generated. Each pipe line therefore has its characteristic sound pattern. When a crack occurs this pattern changes from the no-leak case and the location can be found from direct comparison. The technology has become so advanced that information on dents, buckles, ovality, weld penetration, expansion and pipe line footage can be generated.

The equipment s often simple, consisting of sensor, conditioning, and amplifier circuits and suitable output and recording devices. Such a device developed by British Gas. The range of detection is dependent on the pipe line diameter and the type of pig.

## 3. COMPUTER BASED MONITORING SYSTEMS

It is computer-based systems that the greatest amount of data can be gathered, processed, analyzed, and acted upon in the shortest period of time. Programs can calculate the inventory of the line at any time and compare this with accurate measurements at any section in the system. The effects of pressure and temperature on line dimensions for example, can be calculated to provide an accurate estimate of the mass of fluid in the line. Data from a wide range of instruments can be transmitted by telemetry, radio, or phone links to a central computer which monitors the „health" of line continuously. By changing programs and subroutines, a astound of functions and task can be accomplished very easily and cost effectively.

The many functions that can be performed by computer-based systems include not only leak detection but also:
-pig tracking;
-back tracking of fluid;

-inventory accounting;
-on line flow compensation;
-instrument data and malfunctional checking.

Such systems have very rapid response and have the advantage of multiple inputs being required before leaks are declared. Thus some systems can run for short outage periods with no loss of integrity. They are often complex and costly to install but once the initial capital investment has been made running cost are low. The first section in the detailed review of such systems looks at the phenomena that need to e modeled when a leak occurs.

## 4. PIPE LINE LEAK PHENOMENA

When a leak occurs in a pipe line measured pressure downstream of the leak falls but the pressure at the same location is predicted to rise. The first is not difficult to understand as the line is depressurizing as mass leaves through the lea. The second effect can be explained as follows. The equations predict pressure based on measured flow based on measured pressure. As mass leaves the system through the leak hole, a reduced flow at the downstream end is compared to the inlet flow. This may not have changed and so to balance the system, the equations predict a downstream pressure rise. In physical terms the model thinks to line is „packing" and total system inventory is increasing. There is therefore a divergence between measurement and modeled pressure.

The same is true of flow changes, but here the inlet flow could increase due to lower pipe flow resistance between eater and leak while the section outlet flow will fall as mass leaves through the leak instead of passing through the meter. Thus a real imbalance will result. The model however will show an inconsistency since the pressure comparison will indicate line packing and the flow comparison a line unpacking. If selected pressure and flow imbalance limits are exceeded a leak is declared. The magnitude of the leak is predicted from the flow imbalance and the location from the pressure profile imbalance and the flow leak indicators. The impact of instrument accuracy is important from the leak detections. It is vitally important to good leak sizing and location to have the best pipe line instrumentation possible to minimize uncertainty.

## 5. BACKGROUND PHILOSOPHY OF PIPE LINE MODELING

Real time modeling is a technique that uses the full data gathering capabilities of modern digital systems and the computational power of small computers to give accurate „snapshots" of the pipe line. The whole system is under the control of SCADA package of programs, which poll the data stations on the line, process the data, control the running of the transient pipe line model and activate the alarm and leak location routines. In addition to these basic software modules, more complex systems might include a predictive model to analyze „what if" operating scenarios, provide an optimization routine for least cost operating strategies or include a separate man/machine interface for the model system. The SCADA interface is responsible for acquiring the data from the SCADA system and relating them to the model representation of the line. As a point in the system where two large and independently developed systems join, this can be the source of many problems in the implementation of the real time modeling. Once the measurement data have been obtained, noise altering and plausibility checking can be performed prior to running of the model. The model is the mathematical representation of the pipe line and will include such features as elevation data, diameters, valve and pump locations, changes of direction and the location or cross-over and junctions. The model provides data on the flow conditions within the line at intervals between seconds and minutes, depending on operational needs. Whit the data available from both the measurement system and the pipe line model, the real time applications modules are run. These are the leak detection and location routines in the context of integrity monitoring. He leak detection module functions by computing the difference between the modeled flows and pressures and the measured values at all points where measurements not already used as boundary conditions are located. Because the model accounts for normal transient operations, these differences will be small under normal condition. When a leak is present, the differences become larger since the model system does not account for leakage. When these differences exceed preselected values, a leak alarm is declared. Sophisticated voting schemes which require multiple leak indicators to be in alarm for several consecutive time intervals are used to reduce false alarms while maintaining low thresholds. Often a simple pie line balance of the type discussed earlier is used as a back-up to verify the transient model. Response characteristic are, however, much slower than the real time model.

Once the leak detection module declares a leak, the location routine is activated. The location is calculated from the magnitude and distribution of the leak indicators. As an example, in a straight pipe line with an upstream flow discrepancy and a downstream pressure discrepancy as leak indicators, it is an easy calculation to determine where the leak must be such that the leak flow when added to the modeled flow will produce the additional pressure drop observed at the downstream end. Solutions for pipe networks are more complicated and unique locations do not always exist. This might be the case with parallel looped lines. In this case all the calculated locations should be checked.

The components of real time modeling work together to reduce the large volumes of raw data from the data acquisition system to a much smaller number of parameters and alarm that are more meaningful to pipe line operations. In the case of integrity monitoring, this means leak event that could not be detected by inspection of the measured data can be found and isolated quickly and reliably.

## 6. BASIC PIPE LINE MODELING EQUATIONS

The transient pipe line flow model is the heart of a pipe line modeling system. The model computes the state of the pipe line at each time interval for which data are available. The state of the pipeline is defined as a set of pressures, temperatures, flows, and densities that describe the fluids being transported at all points within the system. These quantities are found as the solution to a set of equations which describe the behavior of the pipe line system. These basic equations are the Continuity equation, the Momentum equation, the Energy equation and an equation of state.

The continuity equation enforces the conservation of mass principle. Simply stated. It requires that the difference in mass flow into and out of section of pipe line is equal to the rate of change of mass within the section. This can be expressed mathematically by the relation:

$$\frac{d(\rho A)}{dt} + \frac{d(\rho A V)}{dx} = 0 \qquad (2)$$

The momentum equation describes the force balance on the fluid within a section of pipe line. It requires that any unbalanced forces result in an acceleration of the fluid element. In mathematical form, this is:

$$\frac{dV}{dt} + V \times \frac{d(V)}{dx} + \frac{1}{\rho} \times \frac{dP}{dx} + g \times \frac{dH}{dx} + \frac{fV|V|}{2 \times D} = 0 \qquad (3)$$

The energy equation states the difference in the energy flow into and out of a section equals the rate of change of energy within the section. This equation is:

$$\frac{dT}{dt} + V \times \frac{d(T)}{dx} + \frac{T}{\rho \times c} \times \frac{dP}{dT} \times \frac{dV}{dx} - \frac{f|V^3|}{2cD} + \frac{4U}{\rho cD}(T - T_g) = 0 \qquad (4)$$

These thee are the basic one dimensional pipe flow equations and are present in one form or another in all transient pipe models. What is needed to solve them, however, is in relation between the pressure, density and temperature for the fluid-an equation of state.

The state equation depends on the type of fluid being modeled, as no one equation fully describes the variety of products that are shipped in pipe line. Some of the forms in use include a bulk modulus type of relation of the form:

$$\rho = \rho_0 \left[ 1 + \frac{P - P_0}{B} + \alpha(T - T_0) \right] \quad (5)$$

This is normally used for liquids that can be regarded as incompressible. The bulk modulus B and the thermal expansion coefficient α can be constant or functions of temperature and/or pressure depending on the application. For light hydrocarbon gases a basic equation such as:

$$P = \rho \times R \times Z \times T \quad (6)$$

Is appropriate, where Z (the compressibility) is a known function of temperature and pressure. For reasonable ranges of temperature and pressure a function of the form:

$$\frac{1}{Z} - 1 \cong \frac{P}{T^y} \quad (7)$$

may be adequate. For conditions where fluids are transported at or near the critical point, a more sophisticated correlation is required to obtain the required accuracy but there is still a large uncertainty in the true density under these operating conditions are they should be avoided wherever possible. Many real time systems have been installed on lines carrying ethylene, butane, propane and LPG products are used the Benedict-Webb-Rubin correlation as modified by Starling with reasonable results. A further complication arises if the product is not uniform throughout the system. This can occur due to batching of fluid or from varying inlet condition. The first is more common where different products are shipped in a common line. The properties are essentially discontinuous across the interface of two fluids, but can be considered as uniform within batches. The basic problem here is to keep tack of the location of the interface. Systems with continuous variation in inlet conditions occur n both liquid and gas systems. The variations can result from mixing of fluid of slightly different composition or large variations in supply conditions.

The governing equation s presented is non-linear partial differential equations which are not suitable for machine computation. They have to be solved by implicit or explicit finite difference techniques or the method seems the most appropriate, as the over two methods could give rise to mathematical instabilities if the wrong time step or distance interval is used.

## 7. SYSTEM DESIGN ASPECTS AND GUIDELINES

The availability o leak alarm uptime depends heavily on the system design and the choice of hardware. Generally the more complex the system, the greater the risk of leak indicator loss, but the more accurate the location of the leaks. Design is therefore a compromise between cost, performance, and reliability. For the three simple alarms of flow imbalance, pressure imbalance, and acoustic alarms between station A and B and C the following components are needed:
-MASS FLOW: 2 flowmeter, 2 pressure sensor, 2 temperature sensor, 2 RTUs, 2 communications links, 1 computer (11 elements);
-PRESSURE: 4 pressure sensors, 4 RTUs, 4 communications links, 1 computer (13 elements);
-ACOUSTIC: 2 acoustic monitors, 2 RTUs, 2 communications links, 1 computer (7 elements);

If combined hybrid alarms of flow/acoustic, flow/pressure, or pressure/acoustic are used then the number of components in the chain is increased.

By summing the component availabilities for each element, an uptime for each alarm can be estimated. From such an analysis, the conclusion can be drawn that the system should be made as simple as possible or instruments should be made as simple as possible or instruments should be installed in duplicate to maximize alarm uptime.

## CONCLUSION

The paper cannot do justice in such a short space, to the complex and diverse subject of leak detection. Such systems have been in operation in many forms all over the world, but it is only recently that environmental as well as economic factors have influenced their development. Instrument selection is critical, as is the need to develop better thermodynamic models, for the next generation of systems to become more reliable and accurate.